# Shock Wave Structure in a Strongly Nonlinear Granular Lattice with Viscous Dissipation

E. B. Herbold[1], V. F. Nesterenko[1,2]

[1]*Department of Mechanical and Aerospace Engineering,*

*University of California at San Diego, La Jolla, California 92093-0411, USA*

[2]*Materials Science and Engineering Program,*

*University of California at San Diego, La Jolla, California 92093-0418, USA*

(Dated: June 26, 2006)

Abstract

The shock wave structure in a one-dimensional lattice (e.g., granular chain) with a power law dependence of force on displacement between particles ($F \propto \delta^n$) with viscous dissipation is considered and compared to the corresponding long wave approximation. A dissipative term depending on the relative velocity between neighboring particles is included in the discrete model to investigate its influence on the shape of steady shock profiles. The critical viscosity coefficient $p_c$ is obtained from the long-wave approximation for arbitrary values of the exponent $n$ and denotes the transition from an oscillatory to a monotonic shock profile in strongly nonlinear systems. The expression for $p_c$ converges to the known equation for the critical viscosity in the weakly nonlinear case. This expression for the critical viscosity is comparable to the value obtained in the numerical analysis of a discrete particle lattice with a Hertzian contact interaction corresponding to $n = 3/2$. An initial disturbance in a discrete system approaches a stationary shock profile after traveling a short distance that is comparable to the width of the leading pulse of a stationary shock front. The shock front width is minimized when the viscosity is equal to its critical value.

PACS numbers: 05.45.Yv, 46.40.Cd, 43.25.+y, 45.70-n

## I. INTRODUCTION

It is well known that one dimensional (1D) and ordered two and three dimensional lattices (granular materials) support compressive strongly nonlinear solitary waves for "normal" nonlinear interaction laws between particles [1]. For example, these waves exist when the force between particles exhibit a power-law dependence on displacement ($F \propto \delta^n$) for $n > 1$ [2–5]. The long wave approximations for power-law materials have exact solutions for solitary waves in non-dissipative particle lattices. When lattices are statically compressed they exhibit a dynamic nonlinear behavior that can be tuned [7] from the strongly to the weakly nonlinear regime. The properties of a solitary wave obtained from the long wave approximation agree very well with the results of experimental studies [1, 6, 7, 8]. However, dissipation significantly attenuates compression pulses in many experimental settings in the strongly and weakly nonlinear regimes and should be included to model real granular systems.

There have been a number of analytical and numerical studies that introduce dissipation into the equations of motion for granular systems. The behavior of a viscous granular gas was investigated [9] using a viscous drag term depending on the velocity of individual particles. Coefficients of restitution were used to investigate inelastic collisions between two particles [10, 11] and viscoelastic interactions were introduced in [12, 13]. Attenuation in a one dimensional lattice is analyzed in [14, 15]. In [16] coupled dashpots representing viscous dissipation depending on the relative motion between atoms were introduced for the investigation of steady shock waves in anharmonic spring-mass systems. The influence of Stokes drag force, which is a function of particle velocities, and damping depending on the relative motion of particles on enveloped solitons in anharmonic discrete lattices was considered in [17].

Recent experimental and numerical work on pulse propagation in lattices immersed in media with different viscosities (air, oil and glycerol) [18] demonstrated that a dissipative term based on the relative velocities between particles affect the wave propagation more significantly than Stokes drag. The mechanism of this phenomenon is related to fluid being expelled from and returning to the elastic contact area during dynamic particle interaction.

The dissipation process influences the behavior of media in the transition area within the shock front and determines the shape of shock profile. For example, in a weakly nonlinear system described by the Korteweg de Vries (KdV) equation, two qualitatively different shock profiles exist depending on the value of the viscosity [19]; oscillatory if below the critical value and monotonic if above. These two profiles also appear experimentally in strongly nonlinear systems using geometrically identical lattices of lead and steel particles under similar loading conditions [1]. The energy dissipation in the strongly nonlinear lead particle lattice is attributed to the significant plastic deformation in the vicinity of the particle contact, which resulted in a monotonic shock profile.

To the best of our knowledge there is no result for the critical viscosity that ensures a transition from an oscillatory to a monotonic shock profile in strongly nonlinear discrete lattices (e.g. granular chains) or in the corresponding long-wave approximations.

This paper studies the effects of this dissipative term on the type of shock wave profiles (i.e. monotonic or oscillatory) in a discrete strongly nonlinear system using numerical analysis. A dissipative term based on the relative velocity between neighboring particles is introduced in the discrete equations and a critical viscosity is found using the long wave approximation. The numerical results are then compared to the analytical treatment of the same system in the context of the long-wave approximation.

## II. EQUATIONS OF MOTION FOR A STRONGLY NONLINEAR DISSIPATIVE DISCRETE LATTICE

In the general case an interaction law between neighboring particles in a nondissipative 1D lattice can be described by a function of their relative displacements $f(u_i - u_{i+1})$, where $u_i$ is the displacement of the $i$-th particle. To account for viscous dissipation, a term based on the relative velocities between particles [16–18] is added to the description

$$m\ddot{u}_i = f(u_{i-1} - u_i) - f(u_i - u_{i+1}) + \mu(\dot{u}_{i-1} - 2\dot{u}_i + \dot{u}_{i+1}) \qquad (1)$$

where $m$ is the mass and $\mu$ is the viscosity coefficient. A compressive solitary wave exists for Eq. (1) in the absence of dissipation and a stationary shock wave exists when dissipation is included for a "normal" interaction law in the continuum approximation of this discrete system [1,20]. The term "normal" refers to an increasing repulsive potential as the distance between particle centers decreases ($f'' > 0$) [1]. In the case of abnormal interactions ($f'' < 0$) solitary and shock rarefaction waves may exist in the frame of long wave approximation. The various properties of these two types of waves are discussed in [1]. The existence of solitary waves in a non-dissipative discrete lattice for "normal" interactions is proven in [20]. As a subset of Eq. (1), particles interacting according to a power-law potential represent a relatively broad class of interactions where exact solutions of long-wave approximations can be found for solitary waves [21]. Numerical analyses for various values of $n$ agree well with the results obtained in the long-wave approximation [15, 22, 23]. The system of equations using a power law potential including a viscous dissipation term is

$$\ddot{u}_i = A_n (u_{i-1} - u_i)^n - A_n (u_i - u_{i+1})^n + p(\dot{u}_{i-1} - 2\dot{u}_i + \dot{u}_{i+1}), \qquad (2)$$

where $A_n$ depends on the geometry of the region of contact and the material properties. The damping coefficient $p$ is defined $p = \mu/m$, where $\mu$ is analogous to the constant used for a dashpot model [16, 17]. The classical Hertzian interaction between perfectly elastic spherical particles is a special case of Eq. (2) when $n = 3/2$. In this case, experimental results agree qualitatively with the long-wave approximation and numerical calculations for a 1D discrete particle lattice made from various types of materials [1,6]: though dissipation was noticeable in all of the experiments.

### III. THE LONG WAVE APPROXIMATION AND CRITICAL VISCOSITY

The long wave approximation is derived from the strongly nonlinear Eq. (2) by assuming that the particle diameter, $a = 2R$, is significantly smaller than the propagating wavelength $L$ so that $\varepsilon \equiv a/L \ll 1$. The expansion in [24] can be used to approximate Eq. (2):

$$u_i = u(x), \qquad u_{i-1} = \exp(-a\partial/\partial x) u(x), \qquad (3)$$
$$u_{i+1} = \exp(a\partial/\partial x) u(x).$$

The result is the long wave equation for strongly nonlinear waves with dissipation [1],

$$u_{tt} = -c_n^2 \left[ (-u_x)^n + \frac{na^2}{24} \left( (n-1)(-u_x)^{n-2} u_{xx}^2 - 2(-u_x)^{n-1} u_{xxx} \right) \right]_x + pa^2 u_{txx} \qquad (4)$$

where $c_n^2 = A_n a^{n+1}$. Terms of the order $O(\varepsilon^6)$ and higher are omitted in the expansions of Eq. (2) as well as the convective derivatives, which is valid for a certain range of wave amplitudes [1]. It should be noted that $c_n$ is not the long-wave sound speed $c_0$. The expression for $c_0$ can be obtained based on the linearization of Eq. (4), $c_0 = c_n \sqrt{n}\, \xi_0^{(n-1)/2}$ [1], where $\xi_0$ is the initial strain due to static compression. Stationary solutions of Eq. (4) without the viscous term have been discussed in [1] and verified numerically in [15, 22, 23, 25-27] for various values of $n$.

We would like to analyze the stationary shock solutions of Eq. (4) $u(x, t) = u(x - Dt)$, where $D$ is the shock wave velocity. This solution satisfies the following, Eq. (5), where the strain is defined as $\xi(x) \equiv -u_x$

$$\frac{D^2}{c_n^2}\xi_x = \left[\xi^n + \frac{na^2}{24}\left((n-1)\xi^{n-2}\xi_x^2 + 2\xi^{n-1}\xi_{xx}\right)\right]_x - \frac{pa^2 D}{c_n^2}\xi_{xx}. \tag{5}$$

The replacement $z = \xi^{(n+1)/2}$ is used to simplify Eq.(5), which can be integrated from $x$ to $\infty$ with the boundary conditions $z(+\infty) = z_0$, $z_x(x = +\infty) = z_{xx}(x = +\infty) = 0$,

$$\frac{D^2}{c_n^2}z^{2/(n+1)} = z^{2n/(n+1)} + \frac{a^2 n}{6(n+1)}z^{(n-1)/(n+1)}z_{xx} - \frac{2pa^2 D}{c_n^2(n+1)}\frac{z_x}{z^{(n-1)/(n+1)}} + C_1. \tag{6}$$

The variable replacements $z = (D/c_n)^{(n+1)/(n-1)}y$ and $x = a\eta\sqrt{n/6(n+1)}$ simplify Eq. (6) to another nondimensional form,

$$y_{\eta\eta} - \bar{p}y^{-2(n-1)/(n+1)}y_\eta + y - y^{-(n-3)/(n+1)} + y^{-(n-1)/(n+1)}C_2 = 0, \tag{7}$$

where $\bar{p}$ represents the dimensionless viscosity:

$$\bar{p} \equiv \frac{2ap}{D}\sqrt{\frac{6}{n(n+1)}}. \tag{8}$$

Equation (7) can be expressed as an equation for the nonlinear oscillator moving in a "potential field" $W(y)$ with a nonlinear "dissipative" term,

$$y_{\eta\eta} = -\frac{\partial W(y)}{\partial y} + \bar{p}y^{-2(n-1)/(n+1)}y_\eta, \tag{9}$$

where the "potential" $W(y)$ is defined

$$W(y) = \frac{1}{2}y^2 - \frac{n+1}{4}y^{4/(n+1)} + C_3 y^{2/(n+1)}. \tag{10}$$

The relations between constants $C_3$, $C_2$ and $C_1$ are

$$C_3 = \frac{n+1}{2}C_2 = \frac{n+1}{2}\left(\frac{c_n}{D}\right)^{2n/(n-1)} C_1. \tag{11}$$

As a representative example of a real system where particles interact according to Hertz law, the potential $W(y)$ is plotted in Fig. 1 and five curves are shown for different values of $C_3$ when $n = 3/2$. All of the curves (1)-(4) in Fig. 1 have a minimum at $y_2$ and maximum at $y_1$. No stationary solitary waves are permitted in the system represented by curve (5) where there is no potential well.

The motion of a nondissipative oscillator with a total energy equal to the maximum of $W(y)$ at $y_1$ corresponds to the solitary wave solution [1]. Nondissipative oscillations corresponding to curve (4) have an initial energy close to the maximum of $W(y)$, which result in a relatively small range of "displacement" $y$. This type of system is related to weakly nonlinear waves that exist in a system having a high initial strain with respect to the dynamic change in strain caused by the wave. Curve (1) corresponds to the strongly nonlinear case of a sonic vacuum where the initial strain is equal to zero and the ratio of the strain in the wave to the initial strain is infinite.

The effective potential energy $W(y)$ has a minimum and maximum if $C_3$ is positive and smaller than some critical value (5/27 in case if $n=3/2$) for $n >1$ [1]. The minimum and maximum values of the potential in the presence of dissipation can be interpreted as the initial and final states of the stationary shock wave. The maximum of $W(y)$ at $y_1$ is related to the initial strain in front of the shock wave and the minima at $y_2$ corresponds to the final equilibrium state. Each pair of $y_1$ and $y_2$ are uniquely defined by the values of $n$ and $C_3$.

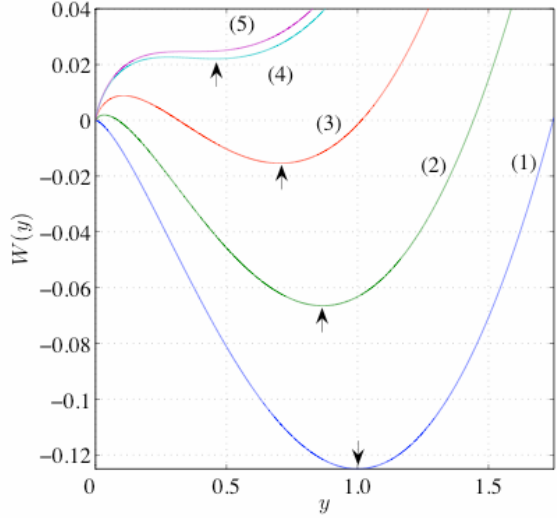

**FIG. 1:** Plot of the potential $W(y)$ for the strongly nonlinear wave equation when $n = 3/2$ (Hertzian potential). Curve (1) $C_3 = 0$, curve (2) $C_3 = 5/81$, curve (3) $C_3 = 10/81$, curve (4) $C_3 = 0.18$, curve (5) $C_3 = 5/27$ (limiting case where there is no minimum or maximum). The arrows indicate the minima of $W(y)$, $y_2$, which is the final state of the shock wave.

It is possible to express $C_3$ in terms of $D$, $c$, and $\xi_0$ for a solitary or shock wave solutions using the condition $\partial W(y)/\partial y\big|_{y=y_1} = 0$,

$$C_3 = \frac{n+1}{2}\left(\frac{c_n}{D}\right)^{2/(n-1)} \xi_0 \left[1 - \left(\frac{c_n}{D}\right)^2 \xi_0^{n-1}\right]. \tag{12}$$

Weakly and strongly nonlinear regimes are determined by the values of $D$ with respect to sound speed $c_0$, which also determines $C_3$. For example, when $D$ approaches the $c_0$ then $C_3$ approaches $5/27$ when $n=3/2$.

The behavior of the strain in a stationary shock profile in the vicinity of the final state of the shock wave can be analyzed by linearizing Eq. (7) assuming,

$$y(\eta) = y_2 + \Psi(\eta), \tag{13}$$

where $\Psi(\eta) \ll y_2$. The arrows in Fig. 1 show $y_2$ corresponding to different values of $C_3$. It is assumed that the transition from an oscillatory to a monotonic shock profile can be identified by the behavior of solutions in the vicinity of the final state represented by $y_2$. Substituting Eq. (13) into Eq. (7) results in the linearized equation,

$$\Psi_{\eta\eta} - \bar{p} y_2^{-2(n-1)/(n+1)} \Psi_\eta + \frac{2}{n+1}\left(n - y_2^{-2(n-1)/(n+1)}\right)\Psi = 0. \tag{14}$$

In the derivation of Eq. (14) $C_3$ was expressed as a function of $y_2$ based on the equation for the derivative of the potential function $\partial W(y)/\partial y|_{y=y_2} = 0$,

$$C_3 = \frac{n+1}{2}\left[y_2^{2/(n+1)} - y_2^{2n/(n+1)}\right]. \tag{15}$$

Equation (14) can be viewed as an equation for a linear oscillator with dissipation. It has the solution

$$\Psi(\eta) = b_1 \exp\left[\left(\frac{\bar{p}}{2} y_2^{-2(n-1)/(n+1)} \pm g(y_2)\right)\eta\right] \tag{16}$$

where $b_1$ is a constant and

$$g(y_2) = \frac{1}{2}\left[\bar{p}^2 y_2^{-4(n-1)/(n+1)} - \frac{8}{n+1}\left(n - y_2^{-2(n-1)/(n+1)}\right)\right]^{1/2}. \tag{17}$$

Imaginary values of $g(y_2)$ correspond to an oscillatory profile and the transition from an oscillatory to a monotonic shock profile occurs when $g(y_2) = 0$. Thus, the critical damping coefficient derives from Eq. (17) and depends on properties of the potential function including power exponent $n$ and position of minimum $y_2$:

$$\bar{p}_c = \sqrt{\frac{8}{n+1}\left(n y_2^{4(n-1)/(n+1)} - y_2^{2(n-1)/(n+1)}\right)}. \tag{18}$$

In the case of a sonic vacuum, $y_2 = 1$ (curve (1) in Fig. 1), which corresponds to $C_2 = 0$ in Eq. (7). Using $y_2 = 1$ in Eq. (18) gives the expression for the dimensionless critical viscosity corresponding to the transition from an oscillatory to a monotonic shock profile in a sonic vacuum for arbitrary $n$,

$$\bar{p}_{c,sv} = \sqrt{8(n-1)/(n+1)}. \tag{19}$$

Combining Eq. (19) with Eq. (8) results in a dimensional form of the critical viscosity:

$$p_{c,sv} = D/a\sqrt{n(n-1)/3}. \tag{20}$$

This critical viscosity depends on the amplitude of the shock wave through its speed $D$, which is related to the final particle velocity and the strain behind the shock. The critical viscosity can be close to zero for very small amplitudes of the shock in a sonic vacuum since $D$ is not restricted by the sound speed.

It is interesting to investigate if the critical viscosity given by Eq. (19) is valid in the entire $y$ domain of the fully nonlinear Eq. (7), though it was derived using Eq. (14) in the vicinity of $y_2$. Figure 2 shows the different types of shock wave propagating in a sonic vacuum for different values of $\bar{p}$ by solving Eq. (7) using Matlab.

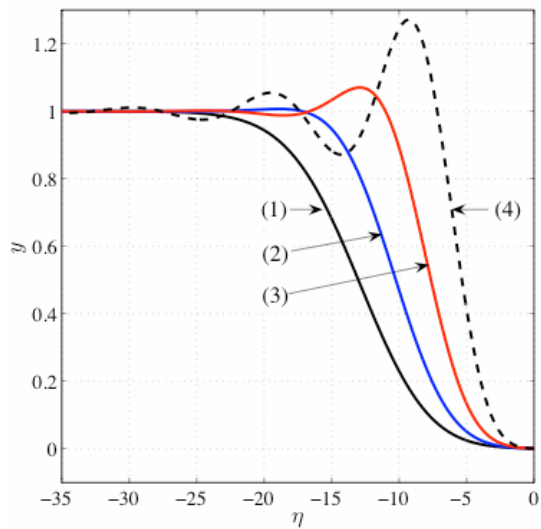

**FIG. 2:** Oscillatory and monotonic shock waves in a "sonic-vacuum". Plot of the numerical solution of Eq. (7) for $n = 3/2$ (Hertzian potential) when $C_3 = 0$ and $\bar{p}_{c,sv} = 1.265$. The curves (1)-(4) in the figure show the solution at different values of $\bar{p}$: curve (1) $\bar{p} = \bar{p}_{c,sv}$, curve (2) $\bar{p} = 0.75\,\bar{p}_{c,sv}$, curve (3) $\bar{p} = 0.5\,\bar{p}_{c,sv}$ and curve (4) $\bar{p} = 0.25\,\bar{p}_{c,sv}$.

The critical value $\bar{p}_{c,sv} = 1.265$ was obtained from Eq. (18) using $y_2 = 1$ and $n = 3/2$ corresponding to a Hertzian interaction. The point $\eta=0$ corresponds to the value of $y$ before the shock wave arrives ($y_1$ corresponding to maximum of $W(y)$ in Figure 1). In the numerical solution to Eq. (7), in the strongly nonlinear regime, an initial displacement of $\Delta y=0.001$ was given to $y$ to start motion from the point corresponding to $y_1$ ($\Delta y$ is equal to 0.1% of the final $y_2$ value). The shock profile corresponding to the critical viscosity $\bar{p} = \bar{p}_{c,sv}$ is shown in Fig. 2(a). It is apparent that the value of $\bar{p}_{c,sv}$ adequately describes the transition from the oscillatory to monotonic profile shown in curves (4)-(1) in Fig. 2.

We conclude that even though the expression for $\bar{p}_{c,sv}$ is derived in the vicinity of the final state of the shock wave, it captures the global transition from oscillatory to monotonic shock profiles in the strongly nonlinear regime corresponding to sonic vacuum.

Remarkably, the reduction of the viscosity resulting in the transition from monotonic to oscillatory shock front does not dramatically reduce the shock onset-width (Fig. 2). The term shock onset width is used here to describe the distance, in $\eta$, from the initial state to the maximum of the first peak of the shock-front. This can be partially explained because the shock onset-width is limited by the half-width of the solitary wave solution in a nondissipative system as $\bar{p} \to 0$. In [1] the equation for the width of a strongly nonlinear solitary wave in a sonic vacuum for particles interacting according to a general power law is,

$$L_s = \pi a/(n-1)\sqrt{n(n+1)/6}, \tag{21}$$

where $a$ is the particle diameter. The width of a strongly nonlinear solitary wave can be expressed in terms of $\eta$ for comparison with the shock onset-width seen in Fig. 2. Dividing this expression by two gives the nondimensional limit of shock onset-width,

$$\bar{L}_s/2 = \pi(n+1)/(2n-2). \tag{22}$$

Substituting $n = 3/2$ into this expression results in an onset-width of 7.8, which is very close to the width shown in curve (4) in Fig. 2 for a relatively small dissipation value $\bar{p} = 0.25\,\bar{p}_{c,sv}$.

It is important to note that the width of the shock front is the distance from the initial state, $y_1$, to the steady final state, $y_2$, which is significantly larger than the onset-width for the oscillatory shock profile. The width of the monotonic shock wave corresponding to $\bar{p}_{c,sv}$ is approximately $7a$, which is about 3 times greater than the solitary wave half-width in a non-dissipative strongly nonlinear sonic vacuum type system (see curve (1) in Fig.(2)). In an over-damped system, where $\bar{p} > \bar{p}_{c,sv}$, the width of the shock front and onset-width are identical and increase with increasing viscosity to values larger than the shock front width when the viscosity is equal to its critical value $\bar{p}_{c,sv}$. For an underdamped system, where $\bar{p} < \bar{p}_{c,sv}$, the shock front width is again longer than in the critically damped case. For example, compare curve (4) in Fig. 2, where $\bar{p} = 0.25\,\bar{p}_{c,sv}$ to curve (1) in Fig. 2, $\bar{p} = \bar{p}_{\mathbf{c,sv}}$.

It is interesting that a long shock front length that is significantly larger than distance between particle centers exists in two distinct cases corresponding to qualitatively different paths to the final state. In the case of weak dissipation the final state is attained through multiple slightly damped oscillations. In the case of a relatively strong dissipation, where $p > p_c$, the final state is attained very slowly without oscillations. As a result the shock front width has a smallest value in a critically damped system at $p = p_c$.

The critical viscosity given by Eq. (18) corresponds to the transition from oscillatory to a monotonic shock profile in the general case including both strongly and weakly nonlinear regimes. It is interesting to compare this prediction of critical viscosity in a weakly nonlinear case with the behavior of solution of Eq. (7). Oscillatory and monotonic profiles of $y$ corresponding to Eq. (7) in the weakly nonlinear potential $W(y)$ (for values of $C_3$ close to 5/27) were obtained using Matlab and are shown in Fig. 3. In the numerical solution of Eq. (7) in the weakly nonlinear regime, a value of $C_3 = 49999/270000$ was used, which is slightly smaller than critical value $C_3 = 5/27$. For this value of $C_3$, the initial and final state in the shock wave are $y_1 = 0.3605474$ and $y_2 = 0.3652323$. An initial displacement of 0.1% of the difference between the starting and final $y$ values ($y = 0.3605521$) was given to $y$ to start a motion from $y_1$.

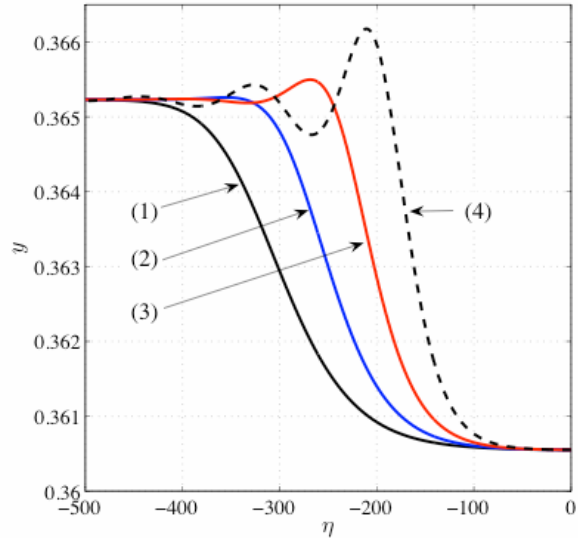

**FIG. 3:** Oscillatory and monotonic shock waves in a weakly nonlinear lattice when $C_3 = 49999/270000$. In this plot of the numerical solution of Eq. (7) the calculated $\bar{p}_{c,w}$ value is 0.074 and curves (1)-(4) in the figure show the solution at $\bar{p}$. Curve (1) corresponds to $\bar{p} = \bar{p}_{c,w}$ curve (2) to $\bar{p} = 0.75\,\bar{p}_{c,w}$ curve (3) to $\bar{p} = 0.5\,\bar{p}_{c,w}$ and curve (4) to $\bar{p} = 0.25\,\bar{p}_{c,w}$.

In this regime the behavior of the solution in the vicinity of $y_2$ is expected to closely match the global behavior of the solution because the change of strain in the wave is small compared to the initial strain of the system. The decrease of viscosity from the critical value $\bar{p}_{c,w}$ taken from Eq. (18) to the smaller values $0.75\,\bar{p}_{c,w}$, $0.5\,\bar{p}_{c,w}$ and $0.25\,\bar{p}_{c,w}$ corresponds to the transition from the monotonic profile curve (1) to the oscillatory shock profiles (2), (3) and (4) in this weakly nonlinear regime.

The equation for the critical viscosity in a system that behaves according to the weakly nonlinear KdV equation is derived in [19]. We can relate the general equation for the critical viscosity Eq. (18) with the known result of the partial weakly nonlinear case using $n = 3/2$ for the Hertzian potential between particles for future comparison with experiments. Also, the KdV equation with dissipation corresponding to Eq. (18) can be easily obtained to apply results of [19] straightforward.

It is assumed that the weakly nonlinear system is initially compressed resulting in initial strain $\xi_0$ and that any traveling wave is a small perturbation: $\xi = \xi_0 + \varepsilon$, $\varepsilon(x,t)/\xi_0 \ll 1$. The initial strain is the second small parameter we need to derive the weakly nonlinear equation in addition to $R/L \ll 1$ required for the long wave approximation, where $L$ is the characteristic wavelength. The resulting wave equation takes the form of the KdV equation including dissipation when written for a wave traveling in one direction,

$$\varepsilon_t - \hat{p}_w \varepsilon_{xx} + c_0 \varepsilon_x + \left(\gamma - \frac{\hat{p}_w^2}{2c_0}\right)\varepsilon_{xxx} + \frac{\sigma}{2c_0}\varepsilon\varepsilon_x = 0, \qquad (23)$$

where $\hat{p}_w \equiv 2R^2 p$ is the viscosity coefficient for a system that permits weakly nonlinear waves (thus the subscript $w$) and has units of dynamic viscosity and

$$\begin{aligned}&\delta_0 = 2R\xi_0\quad,\ c_0^2 = 6AR^2\delta_0^{1/2},\ \sigma = c_0^2 R/\delta_0\\ &c^2 = A(2R)^{5/2},\ \gamma = c_0 R^2/6.\end{aligned} \qquad (24)$$

Note Eq. (23) contains a second order correction to the dissipative term. Neglecting this term is validated because $p_c \to 0$ as $y_2$ and $C_3$ approach their critical values (see curves (4) and (5) in Fig. 1). The equation of critical viscosity for this weakly nonlinear system based on [19] is

$$p_{c,w} = \frac{c_0}{\sqrt{6}R}\sqrt{\frac{D}{c_0}-1}. \tag{25}$$

Equation (25) can also be obtained as the limit of the critical viscosity in the general strongly nonlinear case given by Eq. (18) as $D \to c_0$. For a Hertzian interaction, $n = 3/2$ is placed into Eq. (18) along with the definition of $\bar{p}$ from Eq. (8),

$$p_c = \frac{D}{2\sqrt{2}R}\sqrt{\frac{3}{2}y_2^{4/5} - y_2^{2/5}}. \tag{26}$$

To write this equation in terms of the speed of the shock wave $D$ and initial sound speed $c_0$ we can use the relation between $y_2$ and $y_1$ [28]

$$y_2 = \left[\frac{1}{2}\left(1 - y_1^{2/5} + \sqrt{(1-y_1^{2/5})(1+3y_1^{2/5})}\right)\right]^{5/2}. \tag{27}$$

Using the expressions for $c_0$,

$$c_0^2 = 3/2\,\xi_0^{1/2}c^2, \tag{28}$$

and for $y_1$ in terms of $c_0$ and $D$,

$$y_1 = (2/3)^{5/2}\left(\frac{c_0}{D}\right)^5, \tag{29}$$

and substituting Eqs. (27)-(29) into Eq. (26) obtain

$$p_c = \frac{c_0}{4\sqrt{2}R}\left[\frac{4}{3} - \frac{4}{3}\left(\frac{c_0}{D}\right)^2 + \left(\frac{c_0}{D}\right)^{-2} + \left(1 - 2\left(\frac{c_0}{D}\right)^2\right)\sqrt{\left(\frac{c_0}{D}\right)^{-4} + \frac{4}{3}\left(\frac{c_0}{D}\right)^{-2} - \frac{4}{3}}\right]^{1/2}. \tag{30}$$

When the strongly nonlinear system is precompressed, $D = c_0 + \Delta$ ($\Delta/c_0 \ll 1$), and this small parameter can be used to expand Eq. (30). This expansion results in the critical viscosity in the weakly nonlinear system,

$$p_{c,w} = \sqrt{\frac{c_0\Delta}{6R^2}}. \tag{31}$$

Eq. (25) is recovered from Eq. (31) with the replacement $\Delta = D - c_0$. Thus, Eq. (18) describes the transition from oscillatory to monotonic shock profiles in a general strongly nonlinear case and is consistent with the known equation for the critical viscosity in a weakly nonlinear system.

## IV. NUMERICAL INVESTIGATION OF THE CRITICAL VISCOSITY IN A DISCRETE SYSTEM

Numerical analyses of a discrete particle lattice are presented here for comparison with the results based of long-wave approximation leading to the value of critical viscosity, Eq. (18). There are a few key qualitative differences between the analytical approach in the frame of long wave approximation and numerical calculations. First, the analytical approach assumes a stationary profile with a constant shock wave speed but does not account for the transient development of the wave into its steady state. Additionally, the absence of a restoring force in the discrete system distinguishes it from the continuum in complex ways even in the absence of dissipation [29]. Also, it is important to compare the critical viscosity value derived from the long wave approximation to shock waves in discrete chains because the width of a weakly dissipated shock wave is comparable to the size of the particles; especially for large values of $n$. This may result in a significant difference between the shock wave solutions in the frame of the long wave approximation and the discrete lattice. The numerical analysis will also test how well Eq. (18) and (19) predict the transition from an oscillatory to a monotonic shock profile in a discrete system even though the expressions for $p_c$ rely on the solution in the vicinity of the final state of the shock wave. This numerical investigation will consider a sonic vacuum type system. The solution based on long wave approximation is expected to be a better fit to a behavior of discrete system as it approaches the weakly nonlinear regime.

In the numerical simulation, a shock wave is created by prescribing a velocity $\upsilon_0$ to the first particle of an initially quiescent lattice at $t = 0$. The velocity of the first particle is held constant throughout the calculations. Since a strain in a wave and not the particle displacements are used in the long-wave approximation, we use the discrete displacement solution to find the strains. The equation for the strain on a discrete lattice $\xi$ taken between particles $i$ and $i + 1$ is

$$\xi = (u_i - u_{i+1})/2R. \tag{32}$$

To compare numerical results for not stationary shock wave in discrete lattice with stationary solution in long wave approximation we select the particle velocity in the final state in the shock wave being equal to the velocity of the first particle in numerical calculations. It should be mentioned that final state in the shock wave in sonic vacuum in long wave approximation corresponds to the value of $y = 1$ (see Fig. 1) resulting in the following relations between shock speed $D$, particle velocity $\upsilon_0$ and strain in the final state $\xi_{sh}$:

$$D = c_n^{2/(n+1)} \upsilon_0^{(n-1)/(n+1)} = c_n \xi_{sh}^{(n-1)/2}. \tag{33}$$

The plots for the results of numerical calculations and for the curves obtained in the long wave approximation are presented in nondimensional coordinates $y$ and $\eta$

$$y = (c_n / \upsilon_0) \xi^{(n+1)/2}, \tag{34}$$

$$\eta = -\frac{c_n^{2/(n+1)} \upsilon_0^{(n-1)/(n+1)} t}{2R} \sqrt{6(n+1)/n}. \tag{35}$$

We compare the numerical solution for oscillatory and monotonic shock waves with the long wave approximation using Eq. (19) with $n = 3/2$ for spheres interacting with a Hertzian potential. The values of $p$ in the numerical analysis were taken $p = 0.1 p_c$ and $p = p_c$. Note that $p$ is used instead of $\bar{p}$ because the displacements, $u_i$, are converted to discrete $y$ values after the simulation. However, the values of $p$ in the numerical simulations were calculated to match the $\bar{p}$ values for the comparison with the long-wave approximation. The data in Table I were selected to resemble a real system similar to those found in previous experimental work [7, 18, 29] for the purpose of future verification. The expression for $A$ in the numerical analysis of Eq. (2) assumes a homogeneous particle mass and radius throughout the lattice: $A = E(2R)^{1/2} / (3m(1-v^2))$, where $m = 4/3 \pi R^3 \rho_0$.

**TABLE I: Parameters Used in Numerical Analysis**

|  | Symbol | Units | Value |
|---|---|---|---|
| Young's Modulus | $E$ | [GPa] | 193 |
| Poisson's Ratio | $v$ | - | 0.3 |
| Density | $\rho$ | [kg/m3] | 8000 |
| Particle Radius | $R$ | [m] | 2.38 10$^{-3}$ |
| Mass | $m$ | [kg] | 4.52 10$^{-4}$ |
| Critical Viscosity | $\mu_{cr}$[a] | [Ns/m] | 32.15 |
| Number of Particles | $N$ | - | 1000 |
| Initial Velocity | $\upsilon_0$ | [m/s] | 0.5 |
| Time Step | $\delta t$ | [μs] | 0.875 |

[a] $\mu = m p_c$

The following Figs. 4 to 6 are presented to illustrate both the developing and final state of an oscillatory and monotonic shock wave in a discrete system using 1000 particles with a power law exponent $n = 3/2$ and sonic vacuum initial conditions. The value of $p_c$ used to investigate the behavior of the discrete system is taken directly from the combination of Eq. (19) and Eq. (34) and the values from Table I with 'sonic vacuum' initial conditions.

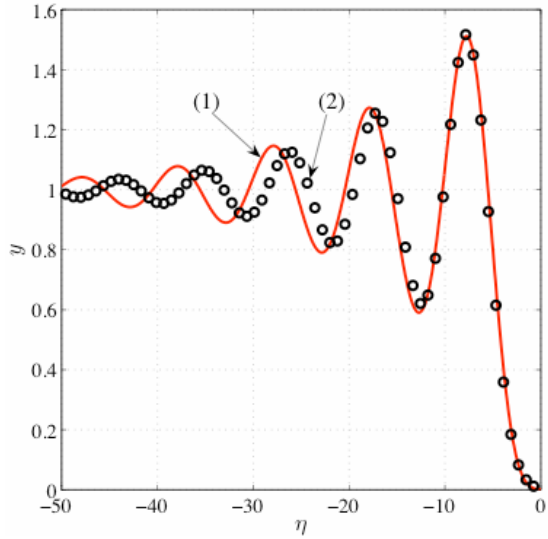

FIG 4: Early development of an oscillatory shock wave in a discrete strongly nonlinear lattice. Curve (1) The numerical solution of the long-wave approximation, Eq. (7), for $\bar{p}$ = 0.1 $\bar{p}_c$. Curve (2) The circles represent the path of $y$ in a discrete particle lattice for the interaction of the 15$^{th}$ and 16$^{th}$ particle from $t = 48.13 \mu s$ (far right) to $t = 102.4 \mu s$ (far left).

Figure 4 depicts the early development of an oscillatory shock wave in a discrete lattice when $p = 0.1 p_c$ from $t = 48.13 \mu s$ to $t = 102.4 \mu s$. The points comprising curve (2) in Fig. 4 are the discrete $y$ values between the 15$^{th}$ and 16$^{th}$ particles. Note the slightly larger amplitude of the first peak and the underdevelopment of the oscillations behind it in comparison to the analytical solution shown in curve (1). Despite this, the qualitative behavior of the shock wave matches well with the long-wave approximation even though it is unsteady.

The early development of a monotonic shock profile at a distance close to the entrance in a discrete lattice is shown in Fig. 5 curve (2) for the critical viscosity $p_c$ found in the long wave approximation. Curve 1 in Fig. 5 represents the stationary profile of a shock wave in the long wave approximation (it is identical to curve (1) in Fig. 2) for comparison to the results for discrete lattice. The $y$ values for the contact between $5^{th}$ and $6^{th}$ particles are shown from $t = 5\mu s$ to $t = 35\mu s$.

In the unsteady state, curve (2) oscillates slightly as it approaches $y = 1$ behind the shock front and the shock onset width is less than that of curve (1). Despite this oscillation it is interesting that even in this unsteady state the shape of the shock profile closely resembles the steady state solution for the long-wave approximation. This shows that the critically damped shock profile in a strongly nonlinear discrete system approaches a stationary state after traveling only a few particles after formation.

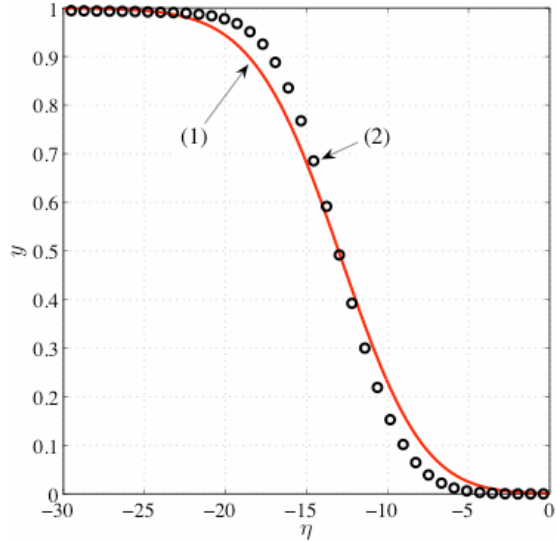

**FIG. 5:** Comparison of early development of a monotonic shock wave in a discrete strongly nonlinear lattice and stationary solution in long wave approximation. Curve (1), the solid line is a plot of the solution of the long-wave approximation, $y(\eta)$, for $\bar{p} = \bar{p}_c$. Curve (2), the circles represent the path of $y$ in a discrete lattice for the interaction of the $5^{th}$ and $6^{th}$ particle from $t = 3.5\mu s$ to $t = 35\mu s$ (traveling from right to left).

In Figs. 4 and 5 the unsteady shock profiles formed close to the entrance were compared to the stationary solution in long-wave approximation. It is interesting to compare them at larger distances from the entrance where shock profiles in the discrete lattice should be closer to a shape corresponding to steady state. We assume that if a profile is not changing qualitatively after traveling through a few hundred particle contacts, it is steady enough for the present discussion. Both oscillatory and monotonic profiles in the long-wave approximation and the discrete lattice at a few distances from entrance are shown in Fig. 6. The stationary shapes of the oscillatory shock waves shown in curves (1) and (2) in Fig. 6(a) are interesting for a number of reasons. The amplitudes of the oscillations behind the leading pulse are slightly larger than in the nonstationary case corresponding to curve (2) in Figure 4.

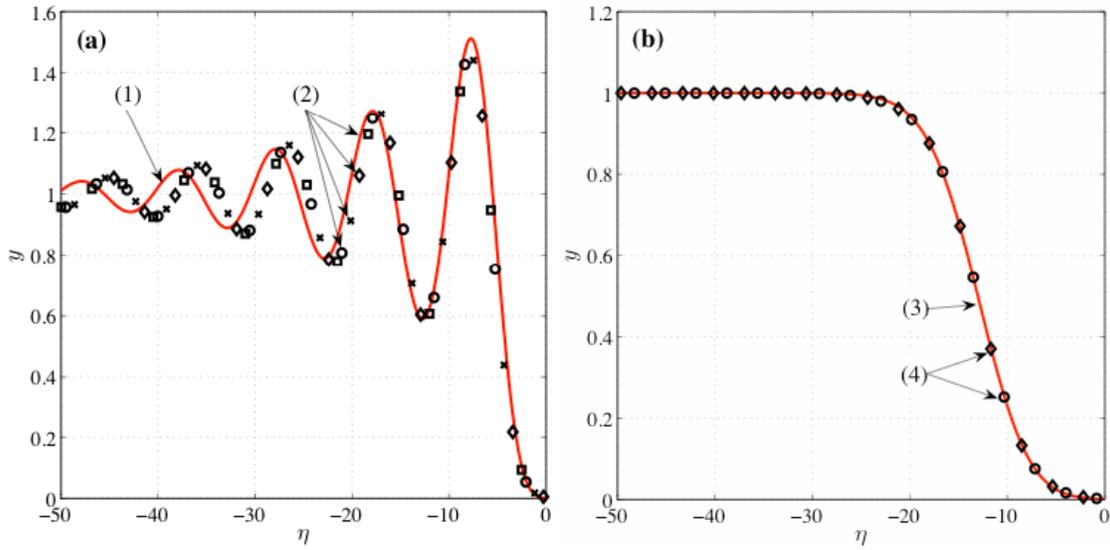

**FIG. 6:** Comparison of oscillatory and monotonic steady shock waves in a discrete strongly nonlinear lattice and the results of the long wave approximation. Curve (1) is a stationary solution of the long wave approximation for an oscillatory shock front; curve (2) is a set of discrete points representing the parameter $y$ related to the strain between particles: □ -- particle contacts 218-234 at time $t \cong 1638$ $\mu s$, ◊ -- particle contacts 454-468 at $t \cong 3278\mu s$, × -- particle contacts 685-701 at $t \cong 4918\mu s$, ○ -- particle contacts 919-934 at $t \cong 6557\mu s$; curve (3) is a stationary solution of the long wave approximation for a monotonic shock front; curve (4) is set of discrete points representing the parameter $y$ between particles for comparison with curve (c): ◊ -- particle contacts 455-470 at $t \cong 3278\mu s$, ○ -- particle contacts 921-937 at $t \cong 6557\mu s$.

Curve (4) in Fig. 6 closely matches the long-wave approximation curve (3) for the monotonic shock wave. The tiny oscillations on the initial stage of shock propagation (they hard to see at the magnification corresponding to Fig. 5(b)) have disappeared as the wave became steady in Figure 6(b), curve (2). The difference in shock-front widths is indistinguishable comparing these two curves. This means that the critical value of viscosity $p_c$ from Eq. (19) captures the transition from oscillatory to monotonic wave profiles very well.

## VI. CONCLUSIONS

The long-wave approximation of a strongly nonlinear system with a power law dependence of force on displacement was extended to include viscous dissipation that depends on the relative velocities of neighboring particles. From this approach an equation for a critical viscosity describing the transition from oscillatory to monotonic shock profiles in a strongly nonlinear regime was derived. This equation naturally includes the weakly nonlinear case. Numerical calculations of the discrete system agreed well with the results of long wave approximation. It should be emphasized that the initial disturbance in a discrete chain approaches a stationary shock regime at a distance comparable to the width of the stationary shock front. This is analogous to the case of a nondissipative system where the initial disturbance forms a single wave or a train of solitary waves after traveling a short distance that is comparable to the width of the solitary wave. The shock front width is minimized when the viscosity is equal to its critical value.

## VII. ACKNOWLEDGEMENTS

The authors wish to acknowledge the support of this work by the U.S. NSF (Grant No. DCMS03013220).

## VIII. REFERENCES

[1] V. F. Nesterenko, Dynamics of Heterogeneous Materials (Springer-Verlag, New York, 2001).
[2] V. Nesterenko, Fizika Goreniya i Vzryva 28, 121 (1992).
[3] V. Nesterenko, in Proceedings of Second International Symposium on Intense Dynamic Loading and Its Effects (Chengdu, China, 1992), pp. 236–240.
[4] V. Nesterenko, in Akustika neodnorodnykh sred (Novosibirsk, 1992), pp. 228–233.
[5] V. Nesterenko, High-Rate Deformation of Heterogeneous Materials (Nauka, Novosibirsk, 1992).
[6] C. Daraio, V. Nesterenko, E. Herbold, and S. Jin, Physical Review E 72, 016603 (2005).
[7] C. Daraio, V. Nesterenko, E. Herbold, and S. Jin, Physical Review E 73, 026610 (2006).
[8] S. Job, F. Melo, A. Sokolow and S. Sen, Physical Review Letters 94, 178002 (2005).
[9] A. Rosas, D. ben-Avraham, and K. Lindenberg, Physical Review E 71, 032301 (2005).
[10] P. Gondret, M. Lance, and L. Petit, Physics of Fluids 14, 643 (2001).
[11] A. Stocchino and M. Guala, Experiments in Fluids 38, 476 (2005).
[12] R. Ramìrez, T. Pöschel, N. V. Brilliantov, and T. Schwager, Physical Review E 60, 4465 (1999).
[13] N. V. Brilliantov, F. Spahn, J.-M. Hertzsch, and T. Pöschel, Physical Review E 53, 5382 (1995).
[14] A. Rosas and K. Lindenberg, Physical Review E 68, 041304 (2003).
[15] M. Manciu, S. Sen, and A. J. Hurd, Physica D 157, 226 (2001).
[16] G. E. Duvall, R. Manvi, and S. C. Lowell, Journal of Applied Physics 40, 3771 (1969).
[17] C. Brunhuber, F. Mertens, and Y. Gaididei, Physical Review E 73, 016614 (2006).
[18] E. Herbold, V. Nesterenko, and C. Daraio (2005), URL http://arxiv.org/abs/cond-mat/0512367.
[19] V. Karpman, Nonlinear Waves in Dispersive Media (Pergamon Press, 1975).
[20] G. Friesecke and J. A. D. Wattis, Communications in Mathematical Physics 161, 391 (1994).
[21] V.F. Nesterenko, Prikl. Mekh. Tekh. Fiz. **5**, 136 (1983) [J. Appl. Mech. Tech. Phys. **5**, 733 (1984)].
[22] E. Hascoët and H. Herrmann, The European Physical Journal B 14, 183 (2000).
[23] A. Rosas and K. Lindenberg, Physical Review E 69, 037601 (2004).
[24] I. A. Kunin, Theory of Elastic Media with Microstructure (Nauka, Moscow (in Russian), 1975).
[25] A. Chatterjee, Physical Review E 59, 5912 (1999).
[26] E. Hinch and S. Saint-Jean, Proceedings of the Royal Society of London 455, 3201 (1999).
[27] C. Arancibia-Bulnes and J. Ruiz-Suàrez, Physica D 168-169, 159 (2002).
[28] C. Coste, E. Falcon, and S. Fauve, Physical Review E 56, 6104 (1997).
[29] V. Nesterenko, C. Daraio, E. Herbold, and S. Jin, Physical Review Letters 95, 158702 (2005).